\documentclass[3p]{elsarticle}

\usepackage{amsfonts, amssymb, array}
\usepackage[fleqn,reqno]{amsmath}
\usepackage{graphics, graphicx, subfigure}
\usepackage{todonotes, comment, soul, float}

\newcommand{\bd}{{\partial}}

\newcommand{\cc}{{\mathbf{c}}}
\newcommand{\CC}{{\mathbb{C}}}

\newcommand{\DDD}{{\boldsymbol{\mathcal D}}}
\newcommand{\eeta}{{\boldsymbol\eta}}

\newcommand{\grad}{{\nabla}}

\newcommand{\llambda}{{\boldsymbol\lambda}}
\newcommand{\nn}{{\mathbf{n}}}

\newcommand{\rr}{{\mathbf{r}}}
\renewcommand{\Re}{{\operatorname{Re}}}
\renewcommand{\Im}{{\operatorname{Im}}}
\newcommand{\RR}{{\mathbb{R}}}
\renewcommand{\ss}{{\mathbf{s}}}

\newcommand{\pp}{{\mathbf{p}}}
\newcommand{\uu}{{\mathbf{u}}}
\newcommand{\UU}{{\mathbf{U}}}

\newcommand{\xx}{{\mathbf{x}}}

\newcommand{\yy}{{\mathbf{y}}}

\def\gap{\hspace*{.2in}}
\newcommand{\pderiv}[2]{\frac{\partial #1}{\partial #2}}

\newcommand{\abs}[1]{\left| #1 \right|}
\newcommand{\Vn}{V_\nn}

\newcommand{\CE}{C_E}
\newcommand{\thL}{$\theta$--$L$}

\newcommand{\bvec}[1]{\mathbf{#1}}
\newcommand {\bq} {\bvec{q}}

\newcommand{\qavg}{\bar{q}}
\newcommand{\pavg}{\bar{p}}
\newcommand{\pup}{p_u}
\newcommand{\pdn}{p_d}
\newcommand{\stress}{{\boldsymbol \sigma}}
\newcommand{\FD}{\bvec{F}_d}
\newcommand{\ex}{ {\bvec{e}}_1}

\newcommand{\anis}{\mathcal{A}}

\newcommand{\diag}{\mathop{\mathrm{diag}}}

\begin{document}
\title{How fluid-mechanical erosion creates anisotropic porous media}


\author[Colgate]{Nicholas J.~Moore}

\author[FSU]{Jake Cherry}

\author[TAMU]{Shang-Huan Chiu}

\author[FSU]{Bryan D.~Quaife}

\address[Colgate]{Colgate University}
\address[FSU]{Florida State University}
\address[TAMU]{Texas A\&M-San Antonio}

\begin{abstract}
Using a Cauchy integral formulation of the boundary integral equations, we simulate the erosion a porous medium comprised of up to 100 solid bodies embedded in a Stokes flow. The grains of the medium are resolved individually and erode under the action of surface shear stress. Through nonlinear feedback with the surrounding flow fields, microscopic changes in grain morphology give way to larger-scale features in the medium such as channelization. The Cauchy-integral formulation and associated quadrature formulas enable us to resolve dense configurations of nearly contacting bodies.
We observe substantial anisotropy to develop over the course of erosion; that is, the configurations that result from erosion generally permit flow in the longitudinal direction more easily than in the transverse direction by up to a factor of six. These results suggest that the erosion of solid material from groundwater flows may contribute to previously observed anisotropy of natural porous media.
\end{abstract}
\maketitle


\section{Introduction}

Flow-induced erosion acts across a range of scales in the natural world, from massive geological structures sculpted by wind or water~\cite{abrams2009growth, perkins2015amplification, mac2020ultra, sharma2022alcove, mac2022morphological}, to mesoscopic patterns formed by surface or internal flows~\cite{berhanu2012shape, bertagni2021hydrodynamic, weady2022anomalous}, and down to granular and porous networks slowly disintegrating in groundwater flows~\cite{chiu2020viscous, szymczak2009wormhole, jager2017channelization, grodzki2019reactive, bizmark2020multiscale, derr2020flow, zareei2022temporal}. The associated nonlinear feedback between changing shapes and the surrounding flows can imprint across all of these scales, affecting large-scale features as well as small-scale ones, such as the microstructure of porous materials. Porous media encountered in nature typically exhibit material {\em anisotropy} in that they permit seepage flow in certain directions more easily than in others. Typical materials are more permeable to flow in the longitudinal (or horizontal) direction than transversely by a factor of 5--20~\cite{bear1988dynamics, anderson2015applied}. Most commonly, this material anisotropy is attributed to the sedimentation process, in which, due to the fluid-structure interaction, non-spherical particles tend to settle with their long axis parallel to the plane of deposition~\cite{bear1988dynamics}. Controlled experiments, however, have not been performed to test this hypothesis, and other mechanisms may be at work. Here, we use highly-accurate numerical simulations to examine an alternative, and possibly complementary, mechanism: namely, that the flow-induced erosion of the medium's solid constituents contributes to its overall anisotropy.

Our method merges highly-efficient and highly-accurate boundary-integral equation (BIE) methods~\cite{baker1986boundary, moore2007evaluation, gray2019boundary} with stable interface evolution methods~\cite{hou-low-she1994, Moore2013} to simulate the erosion of dense suspensions of solid bodies in the Stokes flow regime relevant for groundwater-flow applications~\cite{quaife2018boundary, chiu2020viscous}. Originally inspired by related work in the high-Reynolds-number regime~\cite{Ristroph2012, Moore2013, Huang2015, MooreCPAM2017}, our method is documented, validated, benchmarked in~\cite{quaife2018boundary}, and can simulate the erosion of $O(100)$ solid bodies. The more recent Cauchy reformulation of the BIE and the associated quadrature formulas allow us to resolve points of near contact between bodies, thus enabling high-fidelity simulation of dense suspensions of erodable bodies~\cite{chiu2020viscous}. Though the governing fluid-flow equations (Stokes) are linear, the nonlinear feedback between evolving microstructure and flow gives rise to highly complex and anisotropic configurations.
	
\begin{figure*}
\centering
\includegraphics[width = 0.99 \textwidth]{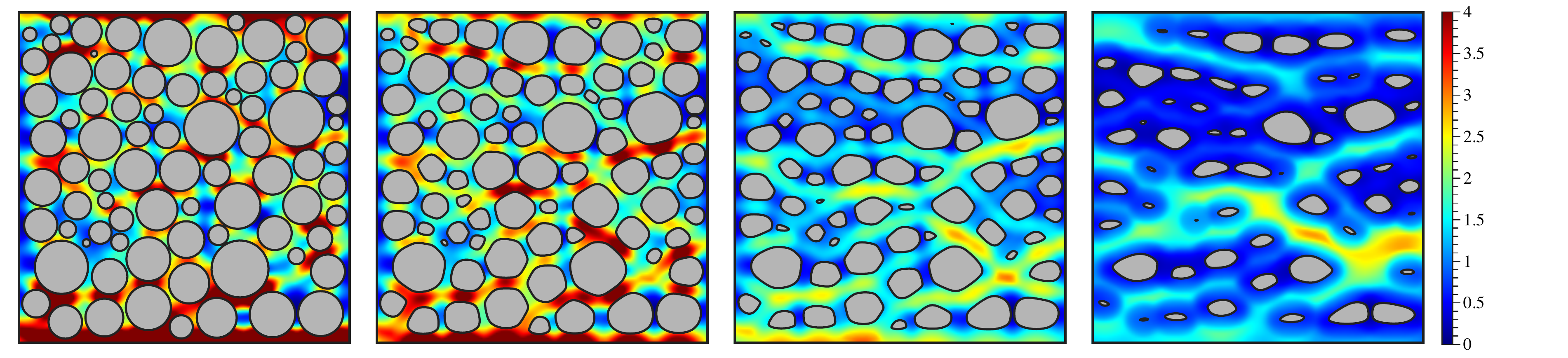}
\caption{The erosion of a porous medium. In this simulation, an initial configuration of 80 circular bodies with random sizes and positions are embedded in a Stokes flow that moves from left to right. The figure shows four snapshots, evenly spaced in time, as the bodies erode in response to the local shear stresses induced by the flow. The local flow speed, shown by color, highlights the appearance of horizontally-aligned channels created by the shape-flow feedback. These channels contribute to the high anisotropy of the medium.
\label{fig1}
}
\end{figure*}

Figure~\ref{fig1} shows an example simulation with 80 circular bodies of randomized sizes and positions immersed in a Stokes flow moving from left to right. 
Color depicts the local speed of the flow intervening between bodies. Over time, individual bodies erode in response to the shear stresses induced on them, and heterogeneous material removal creates visible features both at the level of individual bodies and the larger-scale configuration. For example, horizontally-oriented channels are clearly visible in the third and fourth frames, and these contribute to the overall anisotropy of the medium.


The outline of the paper is as follows. In Section~\ref{sec:formulation} we discuss the governing equations for the fluid flow and interface evolution. In Section~\ref{sec:DLP} we discuss the numerical methods, including the Cauchy formulation of the BIE. In Section~\ref{sec:medium} we explain how to extract porous-medium properties, such permeability, anisotropy, and tortuosity. In Section~\ref{sec:results} we provide results and we conclude in Section~\ref{sec:conclusion}.

\section{Governing Equations}
\label{sec:formulation}

Consider an incompressible, Stokes flow inside a domain $\Omega$ containing $M$ erodable bodies. We take the outer boundary $\Gamma$ to be a slightly smoothed version of the boundary of $[-3,3] \times [-1,1]$. The fluid flow is primarily from left to right, so that the inlet and outlet are located at approximately $x=\pm 3$ (the actual locations are slightly curved versions of these vertical lines due to domain smoothing). The erodable bodies, with boundaries $\gamma_\ell$, $\ell = 1,\ldots,M$, all sit inside of the central region $[-1,1] \times [-1,1]$. The boundary of the fluid domain is thus $\bd \Omega = \Gamma \cup \gamma_1 \cup \cdots \cup \gamma_M$. The empty space to the left and right of $[-1,1] \times [-1,1]$ serve as buffer regions to allow the flow profile imposed at the inlet and outlet to gradually adjust to the presence of the bodies. The equations governing the velocity $\uu = (u,v)$ and pressure $p$ of the fluid consist of the incompressible, Stokes equations coupled to boundary conditions: \begin{equation}
\label{eqn:StokesEq}
  \begin{split}
    \mu \Delta \uu = \grad p, &\hspace{20pt} \xx \in \Omega, \gap 
      &&\mbox{\em conservation of momentum}, \\
    \grad \cdot \uu = 0, &\hspace{20pt} \xx \in \Omega, \gap 
      &&\mbox{\em conservation of mass}, \\
    \uu = \mathbf{0}, &\hspace{20pt} \xx \in \gamma, \gap 
      &&\mbox{\em no slip on the erodable bodies}, \\
    \uu = \UU, &\hspace{20pt} \xx \in \Gamma, \gap 
      &&\mbox{\em prescribed outer wall velocity}.
  \end{split}
\end{equation}
Above, $\UU$ represents the fluid velocity imposed along the outer boundary $\Gamma$, in particular at the inlet and outlet, as well as along the top and bottom walls. In this work, we impose a {\em uniform} flow profile along $\Gamma$, i.e.~$\UU = (U,0)$, although other choices are possible, for example a Poiseuille profile as employed in previous work~\cite{chiu2020viscous, quaife2018boundary}. The advantages of the uniform profile are: (1) it will simplify the calculation of porous-medium properties, such as permeability and anisotropy, that will be described later; and (2), it may more realistically model the flow impinging upon a porous medium. We will allow the imposed flow speed to change with time $U = U(t)$ to enforce, for example, a desired pressure drop across the flow cell. To nondimensionalize the above system, we set the fluid viscosity to unity, $\mu = 1$.

The embedded bodies may erode in response to the shear stresses induced by the intervening fluid flow. Erosion typically occurs over much longer timescales than the fluid flow, permitting a {\em quasi-steady} approximation. In this approximation, the configuration of bodies is held fixed in order to compute the {\em steady} Stokes flow determined by~\eqref{eqn:StokesEq}, and then this flow field determines the stresses acting to erode each body. We employ an erosion law in which the local rate of material loss is linearly proportional to the magnitude of the shear stress $\tau$ acting on the surface~\cite{Ristroph2012, Moore2013, Mitchell2016, MooreCPAM2017, HewettJFS2017, quaife2018boundary, chiu2020viscous}. The material loss gives rise to an inward velocity of the solid surface, $\Vn$, pointing in the direction normal to the surface. The erosion law is thus expressed as
\begin{align}
\Vn = \CE \, \abs{\tau}, 
	&\hspace{20pt} \xx \in \gamma, &&\mbox{\em erosion model}, 
\label{Vn0} \\
\tau = -\mu \left( ( \nabla \uu + \nabla \uu^T) \nn \right) \cdot \ss,
	&\hspace{20pt} \xx \in \gamma, &&\mbox{\em shear stress}.
\label{tau0}
\end{align}
where $\nn$ is the unit normal vector pointing into each body, $\ss$ is the unit tangent vector pointing in the counterclockwise direction, and $\CE$ is a material-dependent erosion constant.



\section{Boundary Integral Equation and Cauchy Integral Formulation}
\label{sec:DLP}
To accurately and efficiently solve the Stokes equations~\eqref{eqn:StokesEq}, we reformulate the system as a boundary integral equation (BIE). A BIE formulation has the advantage that all the unknowns are on the one-dimensional boundaries of the domain. That is, only the boundary of the complex geometry must be discretized which we do with a spectrally accurate Fourier basis. Applying the same approach as our previous works~\cite{chiu2020viscous, quaife2018boundary}, we represent the velocity as the sum of a double-layer potential and a combination of Stokeslets and rotlets~\cite{pow-mir1987}
\begin{align}
  \uu(\xx) = \DDD[\eeta](\xx) + \sum_{\ell=1}^{M}\left(
    S[\llambda_\ell](\xx) + R[\xi_\ell](\xx)\right), 
    \quad \xx \in \Omega,
  \label{eqn:velocity}
\end{align}
where 
\begin{align}
  \DDD[\eeta](\xx) = \frac{1}{\pi}\int_{\bd\Omega} 
  \frac{\rr \cdot \nn}{\rho^2} \frac{\rr \otimes \rr}{\rho^2}
  \eeta(\yy) \, ds_\yy,
  \label{eqn:velocityDLP}
\end{align}
where $\rr = \xx - \yy$ and $\rho = \|\rr\|$. Note that $\bd\Omega$ includes both the eroding bodies and the outer boundary. The Stokeslets and the rotlets are
\begin{align}
  S[\lambda_\ell](\xx) = \frac{1}{4\pi} \left(-\log \rho_\ell +
  \frac{\rr_\ell \otimes \rr_\ell}{\rho_\ell^2} \right) \llambda_\ell,
  \quad
  R[\xi_\ell](\xx) = \frac{\rr_\ell^\perp}{\rho_\ell^2}\xi_\ell,
\end{align}
respectively, where $\rr_\ell = \xx - \cc_\ell$, $\rho_\ell = \|\rr_\ell\|$, and $\cc_\ell$ is a point inside body $\ell$. If the density function, Stokeslets, and rotlets satisfy the second-kind boundary integral equation
\begin{subequations}
  \label{eqn:FredBIE}
  \begin{alignat}{3}
    \UU(\xx) &= -\frac{1}{2}\eeta(\xx) + \DDD[\eeta](\xx) + 
      \sum_{\ell=1}^{M}\left(S[\llambda_\ell](\xx) + R[\xi_\ell](\xx)\right), 
      \quad &&\xx \in \Gamma, \\
    \mathbf{0} &= -\frac{1}{2}\eeta(\xx) + \DDD[\eeta](\xx) + 
      \sum_{\ell=1}^{M}\left(S[\llambda_\ell](\xx) + R[\xi_\ell](\xx)\right), 
      &&\xx \in \gamma_\ell, \: \ell = 1,\ldots,M, \\
    \llambda_\ell &= \int_{\gamma_\ell} \eeta(\yy) \, ds_\yy, 
      &&\ell = 1,\ldots,M, \\
    \xi_\ell &= \int_{\gamma_\ell} (\yy - \cc_\ell)^\perp \cdot 
      \eeta(\yy) \, ds_\yy, &&\ell = 1,\ldots,M,
  \end{alignat}
\end{subequations}
then the representation~\eqref{eqn:velocity} satisfies the Stokes equations with the required boundary conditions~\eqref{eqn:StokesEq}. We solve~\eqref{eqn:FredBIE} by discretizing $\gamma_\ell$ and $\Gamma$ at equispaced collocation points, and then replacing the integrals with quadrature rules. This results in a linear system with a mesh-independent condition number, and it is solved iteratively with GMRES.

Instead of evaluating the Stokes double-layer potential~\eqref{eqn:velocityDLP} as a contour integral in $\RR^2$, we convert the integral to a sum of contour integrals around Jordan curves in $\CC$. We identify $\xx = (x_1,x_2)$ with $x = x_1 + ix_2$ and use similar notation for $\yy$, $\nn$, and $\eeta$. We also interpret $\gamma$, the boundary of the $i^{th}$ grain or the bounding box $\Gamma$, as a Jordan curve in $\CC$. We introduce the functions
\begin{align}
  \tau_1(y) = \eta(y) \overline{n(y)} \Re(n(y)), \quad 
  \tau_2(y) = \eta(y) \overline{n(y)} \Im(n(y)),
\end{align}
and then define the five Cauchy integrals
\begin{align}
  v_1(x) &= \frac{1}{2\pi i} \int_{\gamma} \frac{\Re(\eta(y))}{x-y} \,dy, 
  \quad
  v_2(x) = \frac{1}{2\pi i} \int_{\gamma} \frac{\Im(\eta(y))}{x-y} \, dy, 
  \quad
  v_3(x) = \frac{1}{2\pi i} \int_{\gamma}
  \frac{\Re(\overline{y}\eta(y))}{x-y} \, dy, \\
  v_4(x) &= \frac{1}{2\pi i} \int_{\gamma} \frac{\tau_1(y)}{x-y} \, dy,
  \quad
  v_5(x) = \frac{1}{2\pi i} \int_{\gamma} \frac{\tau_2(y)}{x-y} \, dy.
\end{align}
Then, the first and second components of the Stokes double-layer
potential are
\begin{subequations}
  \begin{align}
    u_1(x) &= -\Re(x)\Re(v_1'(x)) - \Im(x)\Re(v_2'(x)) + 
             \Re(v_3'(x)) + \Re(v_4(x)) \\
    u_2(x) &= +\Re(x)\Im(v_1'(x)) + \Im(x)\Im(v_2'(x)) - 
             \Im(v_3'(x)) + \Re(v_5(x)),
  \end{align}
  \label{eqn:velocityCauchy}
\end{subequations}
respectively. The components of the deformation tensor are
\begin{subequations}
  \begin{align}
    \pderiv{u_1}{x_1} &= -\Re(v'_1(x)) + \Re(v''_3(x)) + \Re(v'_4(x))
                         -\Re(x)\Re(v''_1(x)) - \Im(x)\Re(v''_2(x)) \\
    \pderiv{u_1}{x_2} &= -\Re(v'_2(x)) - \Im(v''_3(x)) - \Im(v'_4(x))
                         +\Re(x)\Im(v''_1(x)) + \Im(x)\Im(v''_2(x)) \\
    \pderiv{u_2}{x_1} &= +\Im(v'_1(x)) - \Im(v''_3(x)) + \Re(v'_5(x))
                         +\Re(x)\Im(v''_1(x)) + \Im(x)\Im(v''_2(x)) \\
    \pderiv{u_2}{x_2} &= +\Im(v'_2(x)) - \Re(v''_3(x)) - \Im(v'_5(x))
                         +\Re(x)\Re(v''_1(x)) + \Im(x)\Re(v''_2(x)) 
  \end{align}
  \label{eqn:deformationCauchy}
\end{subequations}
Having the deformation tensor at hand, the vorticity can be shown to
satisfy
\begin{align}
  \omega(x) = \pderiv{u_2}{x_1} - \pderiv{u_1}{x_2} = 
     \Im(v_1'(x)) + \Re(v_2'(x)) + \Re(v_5'(x)) + \Im(v_4'(x)).
  \label{eqn:vorticityCauchy}
\end{align}
We note that the deformation tensor requires second-order derivatives of
Cauchy integrals, while the vorticity only requires first-order
derivatives. On solid boundaries, the vorticity reduces to the shear stress given in \eqref{tau0} and needed in errosion law \eqref{Vn0} \cite{quaife2018boundary, chiu2020viscous}.

\subsection{Quadrature for Cauchy integrals}
The overall accuracy of our method is determined by the
quadrature rule applied to equation~\eqref{eqn:FredBIE}. Since we have
written the Stokes double-layer potential
velocity~\eqref{eqn:velocityCauchy}, deformation
tensor~\eqref{eqn:deformationCauchy}, and
vorticity~\eqref{eqn:vorticityCauchy} as a sum of Cauchy integrals and
their derivatives, the overall accuracy hinges on the computation of a general Cauchy integral
\begin{align}
  v(x) = \frac{1}{2\pi i} \int_{\gamma} \frac{\eta(y)}{x-y} \, dy.
  \label{eqn:cauchy1}
\end{align}
Here we describe a quadrature formulae that was first used to
approximate analytic functions~\cite{ioa-pap-per1991}, and then extended
to Stokes layer potentials~\cite{bar-wu-vee2015}. The quadrature method
requires the boundary data of the analytic function~\eqref{eqn:cauchy1}
which satisfies the Sokhotski-Plemelj jump relation
\begin{align}
  v^{-}(x_0) = \lim_{\substack{x \rightarrow x_0 \\ x \in \Omega}} \int_{\gamma}
    \frac{\eta(y)}{x-y}\, dy = -\frac{1}{2}\eta(x_0) - 
    \frac{1}{2\pi i} \int_{\gamma} \frac{\eta(y)}{x_0-y} \, dy,
    \quad x_0 \in \gamma,
    \label{eqn:SP}
\end{align}
where the last integral is interpreted in the principal-value sense.
Here we are assuming that $\Omega$ is the bounded region interior to
$\gamma$. Once $v^{-}$ is calculated, $v(x)$ and its derivatives can be
determined by its boundary data alone using the Cauchy Integral Theorem
\begin{subequations}
  \label{eqn:cauchy}
  \begin{alignat}{3}
  \label{eqn:cauchyv}
  v(x) &= \frac{1}{2\pi i}\int_{\gamma} 
    \frac{v^{-}(y)}{y-x} \,dy, \\
  v'(x) &= \frac{1}{2\pi i} \int_{\gamma}
    \frac{v^{-}(y)}{(y-x)^2} \, dy, \\
  v''(x) &= \frac{1}{\pi i} \int_{\gamma}
    \frac{v^{-}(y)}{(y-x)^3} \, dy.
  \end{alignat}
\end{subequations}

The trapezoid rule can be used to approximate this Cauchy integral and
its derivatives. For example, the Cauchy integral~\eqref{eqn:cauchyv}
can be approximated as
\begin{align}
  v(x) \approx \frac{1}{2\pi i} \sum_{j=1}^{N} 
    w_j \frac{v^{-}(y_j)}{y_j - x},
  \label{eqn:trap}
\end{align}
where $y_j$ are equispaced points on $\gamma$, $w_j = L/N$, and $L$ is
the length of $\gamma$. Because the integrand is both periodic and
smooth, given a fixed point $x$, the trapezoid rule achieves spectral
accuracy~\cite{tre-wei2014}. However, for a fixed $N$, the quadrature
error is not bounded uniformly with respect to $x$ because the
derivative of the integrand grows without bound as $x$ approaches
$\gamma$. This error is problematic for many of our simulations since we
allow eroding bodies to be arbitrarily close to one another, and we
often compute the velocity and vorticity at points in the fluid domain
that are close to an eroding body. In contrast to the integrand in the
Cauchy integral~\eqref{eqn:cauchyv}, the integrand in the identity
\begin{align}
  \frac{1}{2\pi i}\int_{\gamma} 
    \frac{v^{-}(y) - v(x)}{y-x} \, dy = 0, \quad x \in \Omega,
  \label{eqn:trap2}
\end{align}
is bounded with respect to $x$, and therefore the error of the trapezoid
rule is bounded with respect to $x$. Applying the trapezoid rule, we
have
\begin{align}
  \frac{1}{2\pi i}\sum_{j=1}^{N} w_{j} 
    \frac{v^{-}(y_j) - v(x)}{y_j - x} \approx 0,
\end{align}
where the error is now uniformly bounded for all $x$. Rearranging, we
have
\begin{align}
  v(x) \approx \left(\frac{1}{2\pi i}\sum_{j=1}^N 
    \frac{v^{-}(y_j)}{y_j - x} w_j \right) \Bigg/
  \left(\frac{1}{2\pi i}\sum_{j=1}^N \frac{1}{y_j - x} w_j \right), 
  \quad x \in \Omega.
  \label{eqn:vBary}
\end{align}
Note that the numerator in~\eqref{eqn:vBary} is identical to
equation~\eqref{eqn:trap}, while the denominator is an approximation of
the analytic function $v(x) = 1$. As $x$ approaches $\gamma$, the errors
in the numerator and denominator grow, however, since the integrand in
equation~\eqref{eqn:trap2} is bounded independent of $x$, the error of
the ratio is also bounded independent of $x$.

The derivatives of the Cauchy integral in~\eqref{eqn:cauchy} can be
approximated with spectral accuracy, uniformly in $x$, using similar quadrature
rules. To summarize, for $x \in \Omega$, 
\begin{align}
  v'(x) &\approx \left(\frac{1}{2\pi i}\sum_{j=1}^{N}
    \frac{v^{-}(y_j) - v(x)}{(y_j-x)^2} w_j \right)
  \Bigg/
  \left(\frac{1}{2\pi i}\sum_{j=1}^{N} \frac{1}{y_j-x} w_j\right), 
  \label{eqn:D1vBary} \\
  v''(x) &\approx \left(\frac{2}{2\pi i}\sum_{j=1}^N 
    \frac{v^{-}_{j} - v(x) - (y_j-x)v'(x)}{(y_j-x)^3}w_j \right)
    \Bigg/
    \left(\frac{1}{2\pi i}\sum_{j=1}^N \frac{1}{y_j-x}w_j\right),
  \label{eqn:D2vBary}
\end{align}
We note that equations~\eqref{eqn:vBary},~\eqref{eqn:D1vBary},
and~\eqref{eqn:D2vBary} all assume that $x \in \Omega$, where $\Omega$
is the bounded region interior to the Jordan curve $\gamma$. However, in
our application, when $\gamma$ is one of the eroding bodies, $x$ is in
the exterior region of the Jordan curve. In this case, slightly
different identities are used, but they all guarantee that the trapezoid
rule achieves spectral accuracy with an error that is independent of
$x$. A complete description of the quadrature rules
for~\eqref{eqn:vBary} and~\eqref{eqn:D1vBary} are described by Barnett,
Wu, and Veerapaneni~\cite{bar-wu-vee2015}, and the quadrature rule
for~\eqref{eqn:D2vBary} is described in our previous
work~\cite{chiu2020viscous}.

\subsection{Interface evolution}

With the flow computed, we extract the vorticity which reduces to shear stress, $\tau$, on solid boundaries. Next, we seek to evolve the boundaries of these erodable bodies. For numerical stability, we modify erosion law \eqref{Vn0} to include a smoothing term that depends on local curvature $\kappa$~\cite{quaife2018boundary}
\begin{align}
  \Vn = \CE \, \abs{\tau} + \epsilon \langle\abs{\tau}\rangle \left(
    \frac{L}{2\pi} \kappa - 1 \right).
\end{align}
The last term is a smoothing term that has strength $\epsilon \ll 1$ and scales with the spatial average of the shear stress $\langle\abs{\tau}\rangle$. $L$ indicates the total arc length of the body, and, inside the parenthesis, the mean curvature is subtracted so that this term preserves area. As such, the only source of material loss is the first term, $\CE \, \abs{\tau}$, representing the shear-dependent erosion law \eqref{Vn0}. In addition, we apply a narrow Gaussian filter to the distribution $\abs{\tau}$ to further improve stability.

Rather than tracking the Cartesian coordinates of each surface, we employ the {\thL} formulation~\cite{hou-low-she1994, Moore2013, MooreCPAM2017, mac2022morphological} by tracking the tangent angle $\theta$ as a function of arc length and the total length $L$ of each body. In this formulation, the curvature-dependent smoothing becomes a linear diffusive term, thus enabling the use of stable, implicit schemes for this stiff term. The remaining nonlinear terms are not stiff and can be treated by explicit time-stepping methods. In particular, we use an exponential integrator for the diffusion term and a Runge-Kutta method for all other terms, both of which are second-order in time~\cite{quaife2018boundary}.

\section{Extracting porous-media properties: permeability, drag, anisotropy, tortuosity}
\label{sec:medium}

With the numerical methods in place, we now discuss how to measure the permeability and other porous-media properties of the configurations generated by fluid-mechanically induced erosion.

\subsection{Longitudinal permeability}
\label{LongPerm}

While the Stokes equations~\eqref{eqn:StokesEq} provides a microscopic description of the detailed flow field $\uu$ penetrating the complex configuration of erodable bodies, a coarse-grained description can be obtained by treating the collection of bodies as a single porous-medium and homogenizing the flow-field through Darcy's law,
\begin{equation}
  \label{eqn:Darcy}
  \bq = - \frac{1}{\mu} \bvec{k} \grad p.
\end{equation}
Here, $p$ and $\mu$ represent the pressure field and fluid viscosity as before, with $\mu = 1$ by the non-dimensionalization. Meanwhile, $\bq = (q_1, q_2)$ represents the {\em specific discharge}, which is the volume of water flowing through a unit cross sectional area of porous media per unit time; $\bq$ relates to the (interstitial) velocity $\uu$, by integrating $\uu$ over a sufficiently small control region and dividing by the total volume (including both fluid and solid) of the region. The parameter $\bvec{k}$ represents the {\em permeability} of the porous medium, which generally takes the form of a rank-2 tensor to permit different propensities to flow in different directions, i.e.~medium anisotropy.  


For simplicity, we assume $\bvec{k}$ to be a diagonal matrix, $\bvec{k}
= \diag(k_{11}, k_{22})$, for the sake of characterizing the
permeability of the porous medium. Because the diagonal components of
$\bvec{k}$ need not be equal, the medium can have different
permeabilities in the longitudinal and transverse directions. Further,
we will assume $\bvec{k}$ to be spatially homogeneous in order to
characterize the medium with a single bulk quantity at any instance in
time. Naturally, the permeability will change over time as the bodies
that comprise the medium disintegrate. Consider first the horizontal
component of~\eqref{eqn:Darcy}
\begin{equation}
  \label{q1}
  q_1(x,y) = -k_{11} \pderiv{p}{x}(x,y).
\end{equation}

By conservations of mass, the average of the horizontal discharge, $q_1$, over any vertical cross-section must equal the uniform flow rate $U$ imposed at the inlet and outlet. That is, for any location $x_0$,
\begin{equation}
\qavg_1 := \frac{1}{2} \int_{-1}^{1} q_1(x_0, y) dy = U.
\end{equation}
Above and henceforth, the overline signifies an average over a vertical cross-section. Similarly, consider the pressure averaged over a vertical cross-section at $x_0$
\begin{equation}
  \pavg(x_0) := \frac{1}{2} \int_{-1}^{1} p(x_0, y) dy.
\end{equation}
In particular, we define the {\em upstream} and {\em downstream} pressures as
\begin{equation}
  \pup = \pavg(-1) \, , \quad \pdn = \pavg(1),
\end{equation}
since $x_0 = -1$ lies immediately upstream of the porous medium and $x_0 = 1$ immediately downstream.

Integrating~\eqref{q1} over the porous-medium domain $[-1, 1] \times [-1,1]$, applying the fundamental theorem of calculus, and rearranging gives
\begin{equation}
  \label{eqn:k11}
  k_{11} = \frac{2U}{\pup - \pdn}.
\end{equation}
This exact formula gives the permeability in terms of the total flux $U$
and measurements of the upstream and downstream pressures.

 

\subsection{Relationship to drag}

The permeability of the medium is directly related to the drag exerted by the collection of bodies. The stress tensor associated with the Stokes equations~\eqref{eqn:StokesEq} is given by $\stress = -p \bvec{I} + \mu \left( \grad \uu + \grad \uu^T \right)$, and the Stokes equations can alternatively be expressed as $\grad \cdot \stress = 0$. Integrating over an arbitrary domain $D$ and applying the divergence theorem gives
\begin{equation}
  \label{divthm}
  0 = \int_{D} \grad \cdot \stress \, dV
  = \int_{\partial D} \stress \, \nn \, ds.
\end{equation}
Consider $D \subseteq [-x_0, x_0] \times [-1, 1]$ to be the subset consisting of the fluid region (i.e.~excluding solid bodies), where $x_0 \ge 1$ will be chosen to include the entire porous region plus some amount of the buffer region. The boundary $\partial D$ consist of all solid-body boundaries $\gamma$, along with an outer boundary. 

The hydrodynamic drag on the collection of bodies is obtained by integrating the surface traction over the boundary $\gamma$. The surface traction on a no-slip boundary is given by 
\begin{equation}
  - \stress \, \nn = p \nn + \tau \ss,
\end{equation}
where the negative sign is a consequence of choosing the normal vector $\nn$ to point out of the fluid region or {\em into} the bodies. The total drag on the collection of bodies is thus
\begin{equation}
  \label{eqn:drag}
  \FD =  - \int_{\gamma} \stress \, \nn \, ds 
  = \int_{\gamma} p \nn + \tau \ss \, ds.
\end{equation}


Projecting~\eqref{divthm} onto the horizontal direction $\ex$, using~\eqref{eqn:drag}, and rearranging gives the exact relationship
\begin{equation}
\label{drag_press}
\FD \cdot \ex = \int_{-1}^{1} \left( -p(x,y) + 2 \mu u_x(x,y) \right) \big|_{x=-x_0}^{x=x_0} dy + 
\int_{-x_0}^{x_0} \mu u_y(x,y) \big|_{y=-1}^{y=1} dx.
\end{equation}
We will now make some simplifying assumptions. First, because the imposed flow profile is uniform $\UU = (U,0)$, slip is permitted along the top and bottom boundaries $y=\pm 1$. Therefore the viscous stress is generally much smaller along these top and bottom surfaces than on the no-slip erodable boundaries. We therefore drop the contribution from the second integral above. Second, if $x_0 > 1$ is chosen a sufficient distance from the erodable bodies, the flow profile approximately matches the uniform profile, implying that the term involving $u_x(\pm x_0, y)$ can be dropped. In addition, the pressure $p(\pm x_0, y)$ approximately matches the downstream and upstream values respectively. With these assumptions,~\eqref{drag_press} simplifies to the approximate form
\begin{equation}
  \FD \cdot \ex \approx 2 (\pup - \pdn).
\end{equation}
Combining with \eqref{eqn:k11} yields a formula relating the longitudinal drag and permeability
\begin{equation}
  \label{perm_drag}
  \frac{1}{k_{11}} \approx \frac{1}{4 U}  \, \FD \cdot \ex
\end{equation}
This formula establishes an important link between the microscopic (Stokes) perspective and the macroscopic (Darcy) perspective, and it gives us a way to test the assumptions involved in coarse-graining the system to extract porous-medium properties. That is, the total drag is an unambiguous quantity that can be computed with high accuracy in our Stokes-based simulations. The extraction of permeability, on the other hand, requires a few key approximations and assumptions, for example that the length-scale of grains is sufficiently small compared to the domain scale so that the flow field can be homogenized. Verifying relationship~\eqref{perm_drag}, as will be done in Section~\ref{sec:results}, will therefore support the idea that the collection of bodies can be treated as a porous medium and meaningful bulk quantities can be extracted.  


\subsection{Transverse permeability and anisotropy}
\label{subsec:anis}

Now consider measuring the transverse permeability $k_{22}$ using Darcy's law. Taking the vertical component of~\eqref{eqn:Darcy} yields
\begin{equation}
  \label{q2}
  q_2(x,y) = -k_{22} \pderiv{p}{y}(x,y).
\end{equation}
This form, however, is not useful if no vertical pressure gradient is imposed, as is the case in our erosion simulations. It is important to recognize that the permeability is a property of the {\em medium}, not the imposed flow. Hence, for a frozen configuration of bodies, it is permissible to alter the imposed flow for the purpose of measuring $k_{22}$ (this altered flow is completely separate from the simulation of the erosion process that generates the configurations). Thus, instead of a horizontal flow in the far-field, we seek to impose a vertical one $\UU = (0,U)$. If this were to be done directly, the outer geometry would need to rotate by 90 degrees about the fixed configuration of bodies. In practice, it is simpler to keep the outer geometry fixed and rotate the inner configuration of bodies, then simply apply the method from Section~\ref{LongPerm} to measure $k_{22}$.

With both the longitudinal and transverse components of permeability computed, we define the anisotropy of the medium as the ratio between the two:
\begin{equation}
  \label{eq:anis}
  \anis = k_{11} / k_{22}.
\end{equation}
Note that the random configuration of circles used to initialize the erosion simulations will have an anisotropy nearly equal to one. As this configuration erodes, it would be expected to permit flow in the longitudinal direction more easily than in the transverse direction, yielding $\anis > 1$.

There are two mechanistic explanations for how erosion can create medium anisotropy. First, the shear stresses could carve each individual body into a more slender form, thus creating anisotropy at the level of individual grains. Second, the shear stressed could preferentially remove certain bodies before others, thus creating large-scale anisotropy. For example, a body positioned in a high-throughput channel might disintegrate relatively quickly, thus opening the channel and creating higher overall anisotropy. We will refer to these two possible mechanisms as {\em shape} anisotropy and {\em configurational} anisotropy, respectively. As an extreme example, a tightly-packed horizontal row of circular bodies would exhibit high configurational anisotropy but no shape anisotropy. Meanwhile, an array of highly eccentric ellipses, all oriented horizontally but positioned randomly, would exhibit high shape anisotropy and low configurational anisotropy. We note that channelization is likely associated with {\em both} types of anisotropy. Certainly the overall configuration must support a channel, and, secondly, the geometry of the individual bodies that outline the channel could control its structure to some degree.

Fortunately, it is possible to devise a test to isolate these two types of anisotropy and therefore determine how much each contributes to the total anisotropy. In particular, for a fixed configuration of partially-eroded bodies, we replace each body with a circle having the same area and center of mass. Since the configurations resulted from the erosion of initially circular bodies, such a replacement does not lead to any overlap between bodies. We then measure the two permeabilities, $k_{11}^{(c)}$ and $ k_{22}^{(c)}$, of this configuration of circular bodies. Since all of the individual shapes are identical and isotropic (i.e.~circles), the resulting anisotropy is entirely due to the relative positions of the bodies, thus allowing us to define the {\em configurational} anisotropy as $\anis_C = k_{11}^{(c)} / k_{22}^{(c)}$. The total anisotropy, meanwhile, is the product of the shape and configurational anisotropy, $\anis = \anis_S \anis_C$, the latter of which we have measured. We can therefore deduce the shape anisotropy through $\anis_S = \anis/\anis_C$.

\subsection{Tortuosity}
%

Like the permeability, the tortuosity provides a macroscopic perspective
of the porous medium \cite{souzy2020velocity}. We define the tortuosity to be the average length
of streamlines passing through the region $[-1,1] \times [-1,1]$. In
particular, for a partially-eroded configuration of bodies with velocity field $\uu$,
we calculate streamlines $\pp(t)$ that satisfy
\begin{align}
  \dot{\pp}(t) = \uu(\pp), \quad \pp(0) = (-1,y_0),
\end{align}
where $y_0 \in (-1,1)$. The streamlines are calculated with a fourth-order Runge-Kutta method. Then, the length of each streamline, which only depends on its initial $y$-coordinate, is
\begin{align}
  \lambda(y_0) = \int_{0}^{t(y_0)} \|\pp(t)\| \,dt.
\end{align}
The integral limit $t(y_0)$ is chosen so that the $x$-coordinate of
$\pp(t(y_0))$ is $+1$. Then, the longitudinal tortuosity is
\begin{align}
  T_1 = \frac{1}{2}\left(\int_{S}u(-1,y)\lambda(y)\,dy \right)
  \Bigg/
  \left(\int_{S}u(-1,y)\,dy \right).
  \label{eqn:tortuosity_x}
\end{align} 
The factor of $1/2$ guarantees that $T_1 \geq 1$, and $T_1 = 1$ if and only if no grains are present. Identical to how we calculate the transverse permeability, we rotate the configuration of eroded bodies by 90 degrees to compute a transverse tortuosity $T_2$.


An alternative and computationally more convenient way to compute the tortuosity is to use an area integral. If there is no reentrant flow in $D = [-1,1] \times [-1,1]$, then the longitudinal tortuosity can be calculated as~\citep{dud-koz-mat2011}
\begin{align}
  T_1 = \left(\int_D \|\uu(\xx)\|\, d\xx \right) \Bigg/
      \left(\int_D u(\xx)\, d\xx \right).
  \label{eqn:tortuosity_x2}
\end{align}
We use a similar area integral to compute $T_2$. We note that there may be slow reentrant regions in our geometries~\cite{chiu2020viscous}, but they are sufficiently small that the difference between equations~\eqref{eqn:tortuosity_x} and~\eqref{eqn:tortuosity_x2} can be neglected.

\section{Results}
\label{sec:results}

With the numerical methods in place, we now present results on how fluid-mechanical erosion alters porous-media properties over time. We first discuss the results from a single simulation and then generalize to statistical analysis of ensembles of simulations.  

\subsection{Single simulation results}
\label{sec:single_sim}

\begin{figure*}
\centering
\includegraphics[width = 0.99 \textwidth]{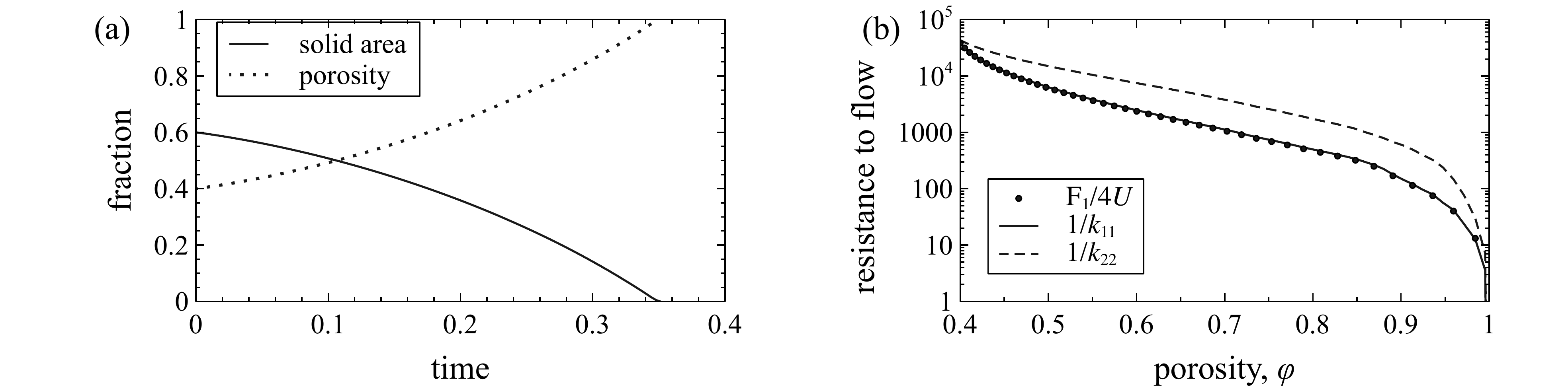}
\caption{A single simulation of porous-medium erosion. The initial configuration consists of 80 circular bodies of random size and position, as seen in Figure~\ref{fig1}. a) As the medium erodes, the fraction of solid-body area decreases, or, equivalently, the porosity increases. b) Resistance to flow can be characterized by either the resistivity (solid) or the cumulative drag (dots), the two of which are related through~\eqref{perm_drag}. Both are seen to decrease as the medium erodes, and relationship~\eqref{perm_drag} is confirmed by the simulation data. The figure also shows the resistivity in transverse direction (dashed).
\label{fig2}
}
\end{figure*}

To begin, we discuss a single simulation of 80 bodies eroding in Stokes flow. The simulation discussed here is the same one shown in Figure~\ref{fig1}. The initial configuration consists of 80 circular bodies having randomized sizes and positions. The material-removal process alters both the shape of individual bodies as well as the overall structure of the pore network transmitting the flow, as can be seen in Figure~\ref{fig1}. Initially, the solid bodies occupy $60\%$ of the area, or the porosity is $\varphi = 0.4$, corresponding to a relatively dense packing; see Figure~\ref{fig1} for a visual. Figure~\ref{fig2}(a) shows how the solid-body area decreases, or equivalently the porosity increases, over time as the bodies erode. The increasing porosity serves as a convenient proxy for dimensionless time that is insensitive to the end conditions setting the strength of the flow (i.e.~whether we specify the pressure drop or the end velocity to be constant in time). We will henceforth use porosity, $\varphi$, on the horizontal axis of many plots to represent increasing time.

As the bodies disintegrate and give way to wider pores, the resistance to flow decreases as seen in Figure~\ref{fig2}(b). The resistance to flow can be quantified in two separate ways, as outlined in Section~\ref{sec:medium}, namely by computing the total drag or by extracting the permeability. As discussed in Section~\ref{sec:medium}, the drag is a microscopic quantity that can be computed with high accuracy and without ambiguity in our Stokes-based simulations. The permeability, on the other hand, relies on a few approximations, but the extraction of such bulk quantities will be more valuable in characterizing the evolving porous medium. As such, it is particularly useful to compare these two perspectives so that the assumptions underlying the coarse-graining process can be assessed. In particular, we aim to test the approximate formula~\eqref{perm_drag}, relating the longitudinal drag $F_1$ to the inverse permeability $1/k_{11}$, also known as the {\em resistivity}. Figure~\ref{fig2}(b) indeed shows close agreement between these two quantities over the entire duration of the simulation, thus confirming the ability to extract medium properties during the erosion process.

In Figure~\ref{fig2}(b) we also show the transverse resistivity, $1/k_{22}$, as it decreases over the course of the simulation. Notice that this transverse resistivity exceeds the longitudinal resistivity by a significant margin, indicating that the configuration provides greater resistance to flow in the vertical direction. This trend fits the intuition that horizontally-aligned channels transmit flow more easily in the longitudinal direction.

\begin{figure*}
\centering
\includegraphics[width = 0.99 \textwidth]{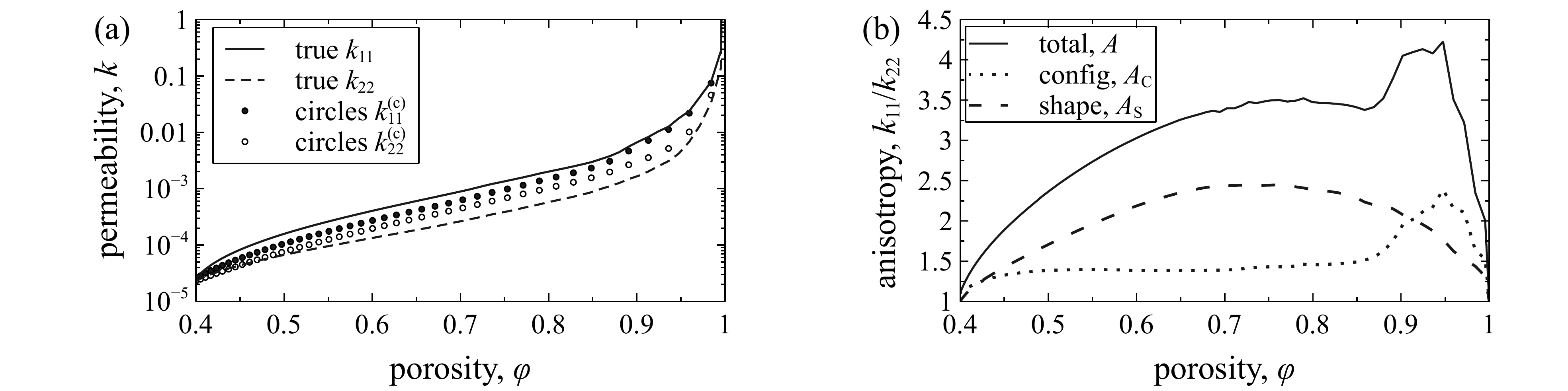}
\caption{
The permeability and anisotropy of the same 80-body simulation featured in Figures~\ref{fig1} and~\ref{fig2}. a) Both the longitudinal (solid) and transverse (dashed) permeability increase over time as the bodies erode. Also shown is the longitudinal and transverse permeabilities of the corresponding configurations of circles, which allows us to extract the configurational component of anisotropy. b) The anisotropy of the medium increases as the bodies erode and preferentially allow flow in the longitudinal direction. The total anisotropy is the product of configurational and shape components. The shape contribution is larger throughout most of the simulation, but near the end of the simulation, the configurational contribution grows large as the channel-structure dominates.
\label{fig3}
}
\end{figure*}

Figure~\ref{fig3}(a) shows more directly how permeability in the longitudinal, $k_{11}$ and transverse, $k_{22}$ directions increase over time as the medium erodes. Once again, the relationship $k_{11} > k_{22}$ indicates a higher propensity for flow in the longitudinal direction. Notice that the permeability increases by nearly five orders of magnitude over the course of the simulation, indicating that erosion substantially alters medium properties.

As given in~\eqref{eq:anis}, the ratio of longitudinal to transverse permeability defines the anisotropy of the medium, $\anis = k_{11} / k_{22}$, which is plotted in Figure~\ref{fig3}(b) over the course of the simulation (solid curve). Early on, the anisotropy is nearly one, as the initial configuration of randomly placed circles has no preferred flow direction. The anisotropy then increases as heterogeneous rates of erosion promote longitudinal flow over transverse flow. The anisotropy peaks at a value of $\anis \approx 4$ later in the simulation. At very late times, the anisotropy approaches unity again as the bodies completely vanish and return the system to a state of no preferred flow direction.

As discussed in Section~\ref{subsec:anis}, the anisotropy of the medium results from the combination of {\em shape} anisotropy, at the individual grain level, and {\em configurational anisotropy}, at the large scale. As detailed in Section~\ref{subsec:anis}, the configurational anisotropy can be extracted through a process in which, at any point in time, the configuration of partially-eroded bodies is replaced with a collection of circles having the same areas and centers of mass. We perform this process and show in Figure~\ref{fig3}(a) the permeabilities, $k_{11}^{(c)}$ and $k_{22}^{(c)}$, of the corresponding circle configurations. As seen in the figure, the permeabilities of the circle configurations always lie in between the values, $k_{11}$ and $k_{22}$, of the true, eroded medium. The {\em configurational} anisotropy is then computed as the permeability ratio of the circle configuration, $\anis_C = k_{11}^{(c)} / k_{22}^{(c)}$, and then the {\em shape} anisotropy can be determined through $\anis_S = \anis/\anis_C$. As seen in Figure~\ref{fig3}(b), the shape anisotropy is the larger factor throughout the majority of the simulation, implying that the shape of individual grains plays a larger role than their relative positions. It is only during the last 20\% of the simulation that the configurational anisotropy grows significantly and surpasses the shape anisotropy. This relative growth of the configurational anisotropy corresponds to the strong channelization seen in the last panel of Figure~\ref{fig1}.

The late emergence of substantial configurational anisotropy may seem
like a spurious effect that is particular to the simulation shown in
Figure~\ref{fig3}. However, statistical analysis of many simulations will show this feature to emerge robustly in simulations featuring a relatively large number of bodies $M \ge 60$.

\begin{figure*}
\centering
\includegraphics[width = 0.99 \textwidth]{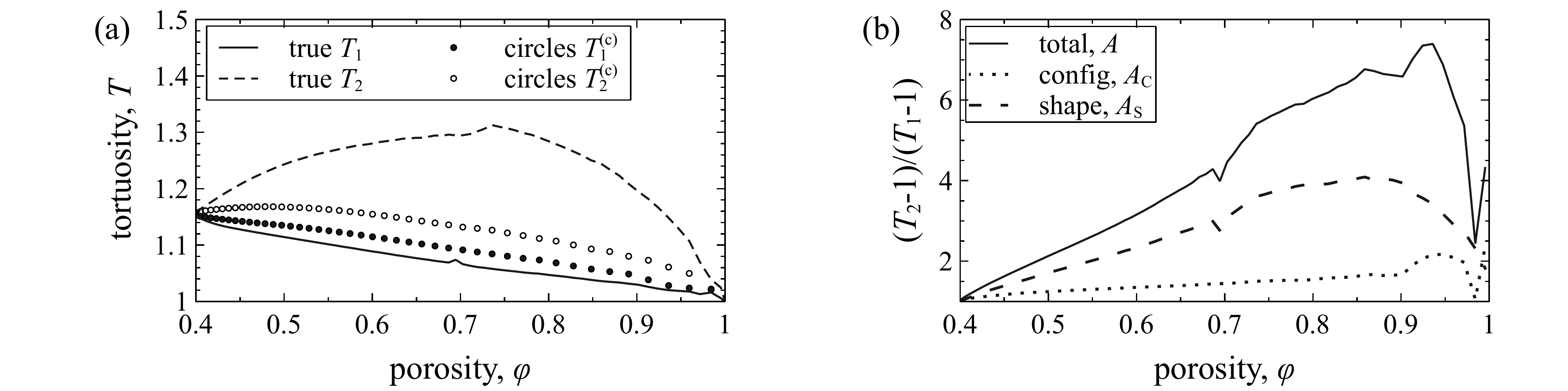}
\caption{
Tortuosity measurements of the same 80-body simulation featured in Figures~\ref{fig1} and~\ref{fig2}. a) As the bodies erode, the transverse tortuosity $T_2$ exceeds the longitudinal component, indicating more convoluted paths for tracers traveling vertically, transverse to the flow. Also shown are the measurements for the configurations of circles. b) The ratio $(T_2-1)/(T_1-1)$ serves as a second proxy for anisotropy. Like anisotropy, this ratio grows large over the course of erosion. It can be decomposed into configurational and shape components in a similar manner.
\label{fig4}
}
\end{figure*}

As a second way to quantify medium anisotropy, we measure the
tortuosity, $T_1$ and $T_2$, in the longitudinal and transverse
directions respectively. Figure~\ref{fig4}(a), shows how both vary over
the course of the simulation. As the bodies erode, $T_2$ grows
significantly larger than $T_1$, indicating that passive tracers must
follow more tortuous paths when traveling transversely compared to longitudinally. The ratio $(T_2-1)/(T_1-1)$ provides
a second measure of anisotropy to complement the permeability-dependent
quantity. We have chosen to subtract one in parenthesis so that both
numerator and denominator both vanish when no bodies are present, as
analogous to resistivity. As seen in Figure~\ref{fig4}(b), the ratio $(T_2-1)/(T_1-1)$ grows significantly during erosion, up to a peak of nearly 8. This ratio can be decomposed into configurational and shape components, exactly as is done for the anisotropy (i.e.~by replacing a collection of eroded bodies with circles of the same areas and centers of mass). Figure~\ref{fig4}(b) shows that the shape component is greater for the majority of the simulation, but near the end of the simulation, the configurational component is competitive.

\subsection{Statistics of ensembles of simulations}

The previous section shows intriguing features to arise from the erosion of a single, random initialization of solid bodies. The observations natural raise the question: are the trends specific to the particular simulation shown or do these features emerge robustly across different initial conditions? To answer this question, we now analyze the same quantities---permeability, tortuosity, and anisotropy---over an ensemble of simulations having different numbers of bodies and different random initializations. We analyze runs having $M$ = 20, 40, 60, 80, and 100 bodies, with at least three different instances of each. This data set represents roughly 2,100 hours of computational time.  

\begin{figure*}
\centering
\includegraphics[width = 0.9 \textwidth]{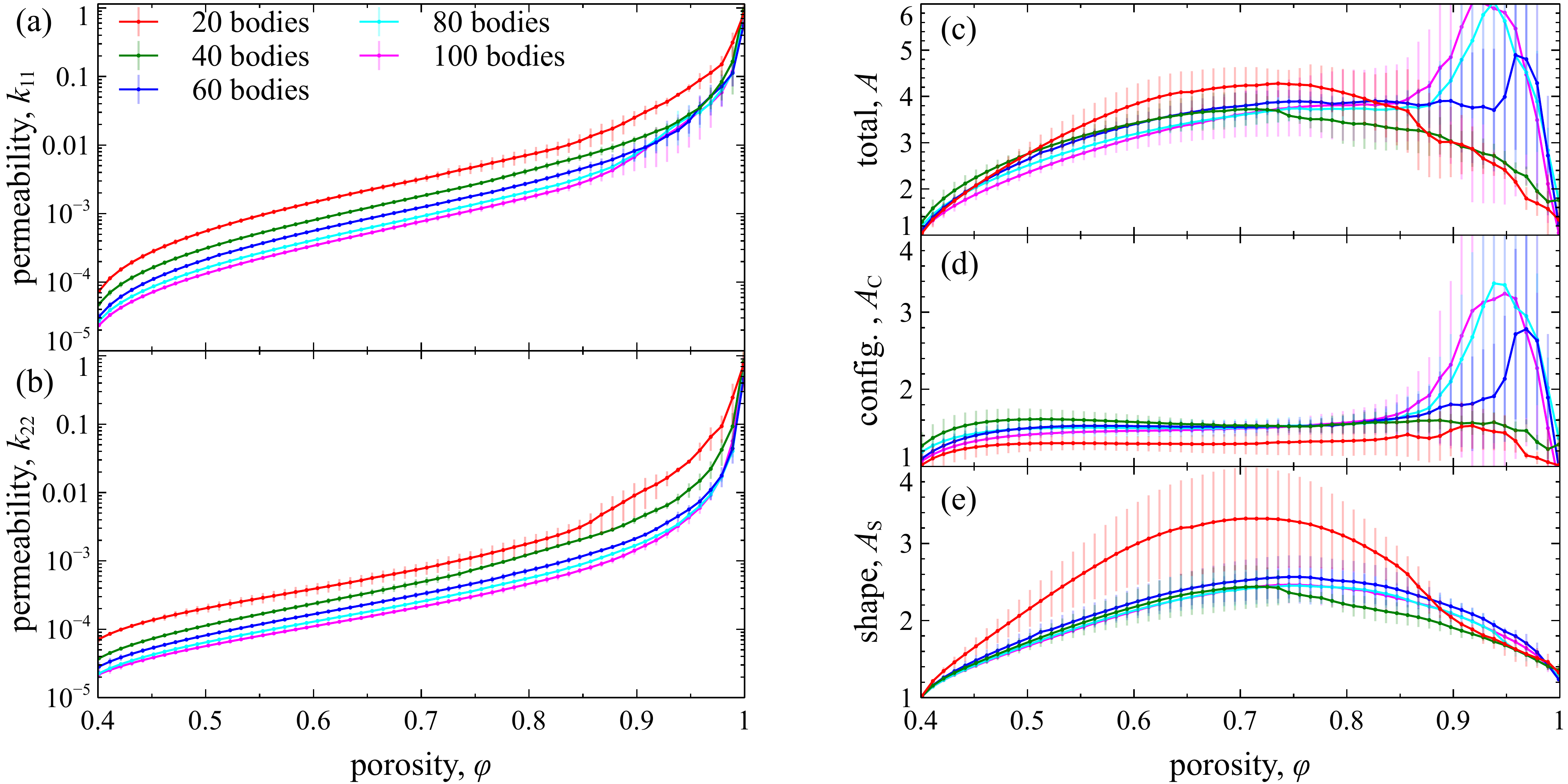}
\caption{Permeability and anisotropy statistics from an ensemble of simulations. (a)--(b) Measurements of the longitudinal (a) and transverse (b) permeability are well separated with respect to the number of bodies, $M$. For each $M$, the mean and standard deviation of permeability are shown against increasing porosity (or time). (c) The same statistics for anisotropy. In all runs, the anisotropy grows substantially and reaches a maximum value of roughly 4--6. Runs with a large number of bodies ($M$ = 60, 80, or 100) show a second surge in anisotropy near the end of the simulation. (d)--(e) The configurational and shape contributions of anisotropy. While the shape anisotropy varies fairly regularly across all runs, the configurational anisotropy shows a distinct peak late in the simulation for runs with large $M$.
\label{fig5}
}
\end{figure*}

Figure~\ref{fig5} shows the permeability and anisotropy measurements for the complete data set. In particular, for a given number of initial bodies ($M$ = 20, 40, 60, 80, or 100), the mean of each quantity (permeability or anisotropy) is plotted against increasing porosity, with the standard deviation across the runs shown by the vertical error bars. Different initial numbers of bodies, $M$, are shown by different colors. As seen in Figures~\ref{fig5}(a)--(b), the trends of increasing permeabilities are well grouped by the number of initial bodies. That is, permeability monotonically decreases with the number of bodies and the error bars shows little overlap between different values of $M$ until very late in the simulation when the bodies have nearly vanished. 

	Figure~\ref{fig5}(c) shows the corresponding statistics of anisotropy, $\anis$. Notice that the anisotropy is not as well separated by $M$. That is, given the error-bars, there is significant overlap in the signal of $\anis$ across different values of $M$. For all values of $M$, the anisotropy initially increases and reaches a peak of roughly $\anis \approx 4$. Interestingly, for the runs featuring a large number of bodies ($M$ = 60, 80, and 100) there is a second rise in anisotropy that occurs late in the simulation and results in a higher peak of roughly $\anis \approx 6$. Thus, the late surge of anisotropy, first observed in Fig.~\ref{fig3}, is not specific to that particular simulation. Rather, it occurs robustly across all simulations as long as the initial number of bodies is sufficiently large, $M \ge 60$.

 	As before, the anisotropy can be decomposed into configuration and shape components, $\anis_C$ and $\anis_S$ respectively, and the statistics of these quantities are shown in Figs.~\ref{fig5}(d)--(e). The shape anisotropy, seen in Fig.~\ref{fig5}(e), takes a fairly regular, parabolic arc --- first increasing due to the shapes carved by erosion and then decreasing as the bodies vanish. This behavior is consistent across all of the values of $M$. The configurational anisotropy (Fig.~\ref{fig5}(d)), however, shows less regular behavior. For runs with a smaller number of bodies ($M$ = 20 and 40) the configurational anisotropy remains relatively small throughout the entire simulation, indicating that the majority of observed anisotropy is due to the individual shapes of bodies. For the runs with a higher body count ($M $ = 60, 80, and 100), though, the configurational anisotropy grows moderately and then surges late in the simulation, as was observed in the single 80-body simulation from Fig.~\ref{fig3}. Since the shape anisotropy is decreasing at this time, the observed second rise in the total anisotropy is due entirely to this surge in the configurational component. For these high-body count runs, the latest stages of erosion are dominated by channelization, which substantially promotes the configurational anisotropy over shape anisotropy. That is, the large-scale arrangement of the bodies has greater effect than the shape of individual bodies.
	
\begin{figure*}
\centering
\includegraphics[width = 0.9 \textwidth]{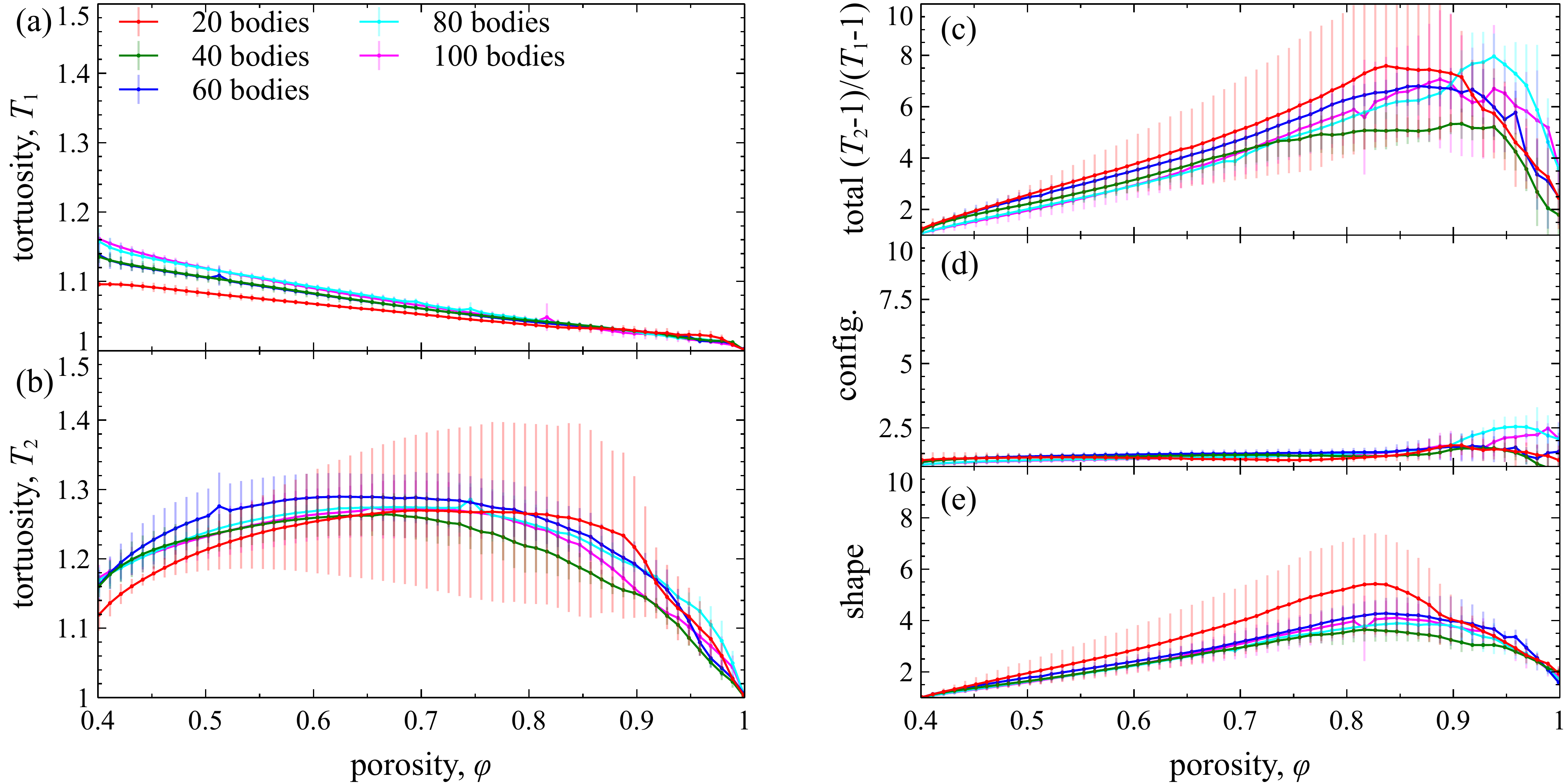}
\caption{
Tortuosity statistics from an ensemble of simulations. (a) The longitudinal tortuosity decreases monotonically with time across all simulations as erosion allows particles to take more direct paths when traveling in the flow direction. (b) The transverse turtuosity initially increases with time as vertically traveling particles are forced to navigate the horizontally aligned shapes carved by erosion. (c)--(e) Statistics of the ratio, $(T_2 -1)/(T_1 -1)$, including its configurational and shape contribution. While the tortuosity ratio reaches a high peak, it is a less sensitive indicator of configurational changes such as channel formation.
\label{fig6}
}
\end{figure*}

We show in Figs.~\ref{fig6}(a)--(b) similar statistical analysis performed on the longitudinal and transverse tortuosity measurements, $T_1$ and $T_2$. Figure~\ref{fig6}(a) shows that, in all simulations, the longitudinal tortuosity decreases with time as erosion allows passive tracers to take more direct paths when traveling in the flow direction. In contrast, Fig.~\ref{fig6}(b) shows the transverse tortuosity increases with time, as vertically-traveling passive tracers must circumvent the horizontally aligned bodies carved by erosion.

As before, the tortuosity ratio, $(T_2 -1)/(T_1 -1)$, provides a second proxy for medium anisotropy to complement the permeability-based definition. Figure~\ref{fig6}(c) shows that the tortuosity ratio behaves similarly across all simulations and all different values of $M$. The ratio increases substantially with erosion, reaching a peak of roughly $(T_2 -1)/(T_1 -1) \approx 8$, before descending during the final stages of erosion. As before the tortuosity ratio can be decomposed into configurational and shape components, as shown in Figures~\ref{fig6}(d)--(e). Here, we see that the shape component is the main contribution and the configurational component remains minimal. This observation fits with the intuition that the main hindrance to a passive tracer should be the shape of individual obstructions, rather than how those obstructions are arranged relative to one another. As such, the tortuosity ratio is a less sensitive indicator of the channelization seen to develop late in the erosion simulations.

\section{Conclusion}
\label{sec:conclusion}

In this paper, we have presented a Cauchy formulation of the boundary integral equations to simulate the fluid-mechanical erosion of many bodies in a Stokes flow. The accuracy and efficiency of the method enables high-fidelity simulations of dense suspensions of $O(100)$ bodies and statistical analysis across an ensemble of different initializations. By linking the governing Stokes equations to Darcy's law, we are able to extract porous-media properties, including permeability, resistivity, tortuosity, and anisotropy, as they evolve over time. Direct comparison between the resistivity and the total drag force confirms such bulk properties are extracted with high accuracy.

The ratio of longitudinal to transverse permeability provides our main diagnostic of medium anisotropy. Measurements indicate that the anisotropy grows substantially as a result of fluid-mechanical erosion, reaching a peak of roughly six in most simulations, before the bodies completely vanish and return the system to an isotropic state. The anisotropy can be further decomposed into a configurational component, due only to the relative positions of bodies, and a shape component, due to the detailed geometry of each body. Statistical analysis across a large number of simulations reveals that if the number of bodies is sufficiently large, $M \ge 60$, the configurational anisotropy surges near the final stages of erosion and surpasses the shape anisotropy as the primary contribution. This surge in configurational anisotropy is associated with the formation of visual channels that transmit a large portion of the flow. The tortuosity provides a second metric and also reveals strong anisotropy to develop across all simulations. In the future, we hope to extend the methodology to more complex scenarios of thermal convective flows in porous media~\cite{McCurdy2019}, erodable porous structures formed by precipitation reactions~\cite{eastham2020multiphase}, and first arrival statistics of Brownian particles~\cite{che-lin-her-qua2021}. 

\section*{Acknowledgements}
N.J.M., J.C., and B.D.Q. recognize the support of NSF Grant DMS-2012560

\bibliographystyle{plain}

\begin{thebibliography}{10}

\bibitem{abrams2009growth}
D.M. Abrams, A.E. Lobkovsky, A.P. Petroff, K.M. Straub, B.~McElroy, D.C.
  Mohrig, A.~Kudrolli, and D.H. Rothman.
\newblock Growth laws for channel networks incised by groundwater flow.
\newblock {\em Nature Geoscience}, 2(3):193, 2009.

\bibitem{anderson2015applied}
M.P. Anderson, W.W. Woessner, and R.J. Hunt.
\newblock {\em Applied groundwater modeling: simulation of flow and advective
  transport}.
\newblock Academic press, 2015.

\bibitem{baker1986boundary}
G.R. Baker and M.J. Shelley.
\newblock Boundary integral techniques for multi-connected domains.
\newblock {\em Journal of Computational Physics}, 64(1):112--132, 1986.

\bibitem{bar-wu-vee2015}
Alex Barnett, Bowei Wu, and Shravan Veerapaneni.
\newblock {Spectrally-Accurate Quadratures for Evaluation of Layer Potentials
  Close to the Boundary for the 2D Stokes and Laplace Equations}.
\newblock {\em SIAM Journal on Scientific Computing}, 37(4):B519--B542, 2015.

\bibitem{bear1988dynamics}
J.~Bear.
\newblock {\em Dynamics of fluids in porous media}.
\newblock Courier Corporation, 1988.

\bibitem{berhanu2012shape}
M.~Berhanu, A.~Petroff, O.~Devauchelle, A.~Kudrolli, and D.H. Rothman.
\newblock Shape and dynamics of seepage erosion in a horizontal granular bed.
\newblock {\em Physical Review E}, 86(4):041304, 2012.

\bibitem{bertagni2021hydrodynamic}
M.B. Bertagni and C.~Camporeale.
\newblock The hydrodynamic genesis of linear karren patterns.
\newblock {\em Journal of Fluid Mechanics}, 913, 2021.

\bibitem{bizmark2020multiscale}
N.~Bizmark, J.~Schneider, R.D. Priestley, and S.S. Datta.
\newblock Multiscale dynamics of colloidal deposition and erosion in porous
  media.
\newblock {\em Science advances}, 6(46):eabc2530, 2020.

\bibitem{che-lin-her-qua2021}
Jake Cherry, Alan~E. Lindsay, Adri\'an~Navarro Hern\'andez, and Bryan Quaife.
\newblock {Trapping of Planar Brownian Motion, Full First Passage Time
  Distributions by Kinetic Monte-Carlo, Asymptotic and Boundary Integral
  Equations}.
\newblock {\em arxiv}, 2112.06842, 2021.

\bibitem{chiu2020viscous}
S.H. Chiu, M.N.J. Moore, and B.~Quaife.
\newblock Viscous transport in eroding porous media.
\newblock {\em Journal of Fluid Mechanics}, 893, 2020.

\bibitem{derr2020flow}
N.J. Derr, D.C. Fronk, C.A. Weber, A.~Mahadevan, C.H. Rycroft, and
  L.~Mahadevan.
\newblock Flow-driven branching in a frangible porous medium.
\newblock {\em Physical review letters}, 125(15):158002, 2020.

\bibitem{dud-koz-mat2011}
Artur Duda, Zbigniew Koza, and Maciej Matyka.
\newblock Hydraulic tortuosity in arbitrary porous media flow.
\newblock {\em Physical Review E}, 84:036319, 2011.

\bibitem{eastham2020multiphase}
P.S. Eastham, M.N.J. Moore, N.G. Cogan, Q.~Wang, and O.~Steinbock.
\newblock Multiphase modelling of precipitation-induced membrane formation.
\newblock {\em Journal of Fluid Mechanics}, 888, 2020.

\bibitem{gray2019boundary}
L.J. Gray, J.~Jakowski, M.N.J. Moore, and W.~Ye.
\newblock Boundary integral analysis for non-homogeneous, incompressible stokes
  flows.
\newblock {\em Advances in Computational Mathematics}, 45(3):1729--1734, 2019.

\bibitem{grodzki2019reactive}
Piotr Grodzki and Piotr Szymczak.
\newblock {Reactive-infiltration instability in radial geometry: From
  dissolution fingers to star patterns}.
\newblock {\em Physical Review E}, 100(3):033108, 2019.

\bibitem{HewettJFS2017}
J.N. Hewett and M.~Sellier.
\newblock Evolution of an eroding cylinder in single and lattice arrangements.
\newblock {\em J. Fluid Struct.}, 70:295--313, 2017.

\bibitem{hou-low-she1994}
T.Y. Hou, J.S. Lowengrub, and M.J. Shelley.
\newblock {Removing the Stiffness for Interfacial Flows with Surface Tension}.
\newblock {\em Journal of Computational Physics}, 114:312--338, 1994.

\bibitem{Huang2015}
J.M. Huang, M.N.J. Moore, and L.~Ristroph.
\newblock Shape dynamics and scaling laws for a body dissolving in fluid flow.
\newblock {\em J. Fluid Mech.}, 765:R3, 2015.

\bibitem{mac2022morphological}
J.M. Huang and N.J. Moore.
\newblock Morphological attractors in natural convective dissolution.
\newblock {\em Physical Review Letters}, 128(2):024501, 2022.

\bibitem{mac2020ultra}
J.M. Huang, J.~Tong, M.~Shelley, and L.~Ristroph.
\newblock Ultra-sharp pinnacles sculpted by natural convective dissolution.
\newblock {\em Proceedings of the National Academy of Sciences},
  117(38):23339--23344, 2020.

\bibitem{ioa-pap-per1991}
N.~I. Ioakimidis, K.~E. Papadakis, and E.~A. Perdios.
\newblock {Numerical Evaluations of Analytic Functions by Cauchy's Theorem}.
\newblock {\em BIT Numerical Mathematics}, 31(2):276--285, 1991.

\bibitem{jager2017channelization}
R.~J{\"a}ger, M.~Mendoza, and H.J. Herrmann.
\newblock Channelization in porous media driven by erosion and deposition.
\newblock {\em Physical Review E}, 95(1):013110, 2017.

\bibitem{McCurdy2019}
M.~McCurdy, N.~Moore, and X.~Wang.
\newblock Convection in a coupled free flow-porous media system.
\newblock {\em SIAM Journal on Applied Mathematics}, 79(6):2313--2339, 2019.

\bibitem{Mitchell2016}
W.H. Mitchell and S.E. Spagnolie.
\newblock A generalized traction integral equation for {S}tokes flow, with
  applications to near-wall particle mobility and viscous erosion.
\newblock {\em J. Comput. Phys.}, 2016.

\bibitem{MooreCPAM2017}
M.N.J. Moore.
\newblock {Riemann-Hilbert Problems for the Shapes Formed by Bodies Dissolving,
  Melting, and Eroding in Fluid Flows}.
\newblock {\em Comm. Pure Appl. Math.}, 2017.

\bibitem{moore2007evaluation}
M.N.J. Moore, L.J. Gray, and T.~Kaplan.
\newblock Evaluation of supersingular integrals: second-order boundary
  derivatives.
\newblock {\em Int. J. Numer. Meth. Eng.}, 69(9):1930--1947, 2007.

\bibitem{Moore2013}
M.N.J. Moore, L.~Ristroph, S.~Childress, J.~Zhang, and M.J. Shelley.
\newblock Self-similar evolution of a body eroding in a fluid flow.
\newblock {\em Phys. Fluids}, 25(11):116602, 2013.

\bibitem{perkins2015amplification}
J.P. Perkins, N.J. Finnegan, and S.L. De~Silva.
\newblock Amplification of bedrock canyon incision by wind.
\newblock {\em Nature Geoscience}, 8(4):305, 2015.

\bibitem{pow-mir1987}
H.~Power and G.~Miranda.
\newblock {Second kind integral equation formulation of Stokes' flows past a
  particle of arbitrary shape}.
\newblock {\em SIAM Journal on Applied Mathematics}, 47(4):689--698, 1987.

\bibitem{quaife2018boundary}
B.~Quaife and M.N.J. Moore.
\newblock A boundary-integral framework to simulate viscous erosion of a porous
  medium.
\newblock {\em Journal of Computational Physics}, 375:1--21, 2018.

\bibitem{Ristroph2012}
L.~Ristroph, M.N.J. Moore, S.~Childress, M.J. Shelley, and J.~Zhang.
\newblock Sculpting of an erodible body by flowing water.
\newblock {\em P. Natl. Acad. Sci. USA}, 109(48):19606--19609, 2012.

\bibitem{sharma2022alcove}
R.S. Sharma, M.~Berhanu, and A.~Kudrolli.
\newblock Alcove formation in dissolving cliffs driven by density inversion
  instability.
\newblock {\em Physics of Fluids}, 34(5):054118, 2022.

\bibitem{souzy2020velocity}
M.~Souzy, H.~Lhuissier, Y.~M{\'e}heust, T.~Le~Borgne, and B.~Metzger.
\newblock Velocity distributions, dispersion and stretching in
  three-dimensional porous media.
\newblock {\em Journal of Fluid Mechanics}, 891, 2020.

\bibitem{szymczak2009wormhole}
P.~Szymczak and A.J.C. Ladd.
\newblock Wormhole formation in dissolving fractures.
\newblock {\em Journal of Geophysical Research: Solid Earth}, 114(B6), 2009.

\bibitem{tre-wei2014}
Lloyd~N. Trefethen and J.~A.~C. Weideman.
\newblock {The Exponentially Convergent Trapezoidal Rule}.
\newblock {\em SIAM Review}, 56(3):385--458, 2014.

\bibitem{weady2022anomalous}
S.~Weady, J.~Tong, A.~Zidovska, and L.~Ristroph.
\newblock Anomalous convective flows carve pinnacles and scallops in melting
  ice.
\newblock {\em Physical Review Letters}, 128(4):044502, 2022.

\bibitem{zareei2022temporal}
A.~Zareei, D.~Pan, and A.~Amir.
\newblock {Temporal Evolution of Erosion in Pore Networks: From Homogenization
  to Instability}.
\newblock {\em Physical Review Letters}, 128(23):234501, 2022.

\end{thebibliography}

\end{document}